\documentstyle{ichep}
\input psfig
\newcommand{\met}{\mbox{$\not\!\!E_T$}}
\newcommand{\ipb}{pb$^{-1}$}

\newcommand{\tbt}{$t\overline{t}$}
\newcommand{\bbb}{$b\overline{b}$}

\newcommand{\et}{$E_T$}
\newcommand{\pet}{$p_t$}
\newcommand{\beq}{\begin{equation}}
\newcommand{\eeq}{\end{equation}}
\newcommand{\gsq}{GeV/c$^2$}
\newcommand{\g}{GeV}
\newcommand{\gc}{GeV/c}
\begin{document}

\title{Search for top in D\O\ using the electron + jets channel
with soft $\mu$ tagging}

\author{Rajendran Raja}

\affil{Fermi National Accelerator Laboratory,\\
Batavia, Illinois 60510}

\collab{For the D\O\ Collaboration}

\abstract{We present preliminary results for the search for the top quark in
D\O\ in the electron + jets channel where one of the $b$ quark jets is tagged
by means of a  soft muon, using  13.5 \ipb\ of data. Standard model decay modes
for the top quark are assumed. \cabs  We present the resulting top cross
section
and error as a function of  top mass using this  channel combined with the
dilepton channel and  the untagged lepton + jets  channel presented elsewhere
in this session. At present, no significant signal for top quark production
can be established.}

\twocolumn[\maketitle]

\section{Introduction}
In the standard model, each top quark decays predominantly to a W boson and a
b
quark. Each \tbt\ pair in an event will thus be accompanied by a \bbb\ pair. If
we assume that each $b$ quark decays  semi-leptonically $\sim$ 10 \% of the
time
into a muon  and likewise for the $c$ quark resulting from the $b$ quark
decay,
$\sim 44 \%$ of the \tbt\ events will have a soft muon. D\O\ has a muon
detection system \cite{det} that is characterized by nearly 4$\pi$ in solid
angle coverage, containing 12-18 interaction lengths of absorber and a
relatively small decay volume in the central tracker. This system is capable of
detecting these muons (the average \pet\ of such muons from a 160 \gsq\ top
quark is 17 \gc\ )with an efficiency such that $\sim 20 \% $ of the \tbt\
events
will have a detected soft muon tag. Because the conventional W + jets
background
to the lepton + jets channel is expected to be much less rich in b quarks, it
is
possible to employ looser cuts in event selection as a result of demanding the
lepton tag.

The results of  top searches employing dilepton channels and lepton + jets
channels without tagging the b quark have been reported \cite{steve,serban}
in this session. We report here the top production cross section and error
combining the results of all these channels. The summary of these three papers
is also given in the plenary session \cite{grannis}.
\section{Estimation of backgrounds}
\begin{figure}
\centerline{\psfig{figure=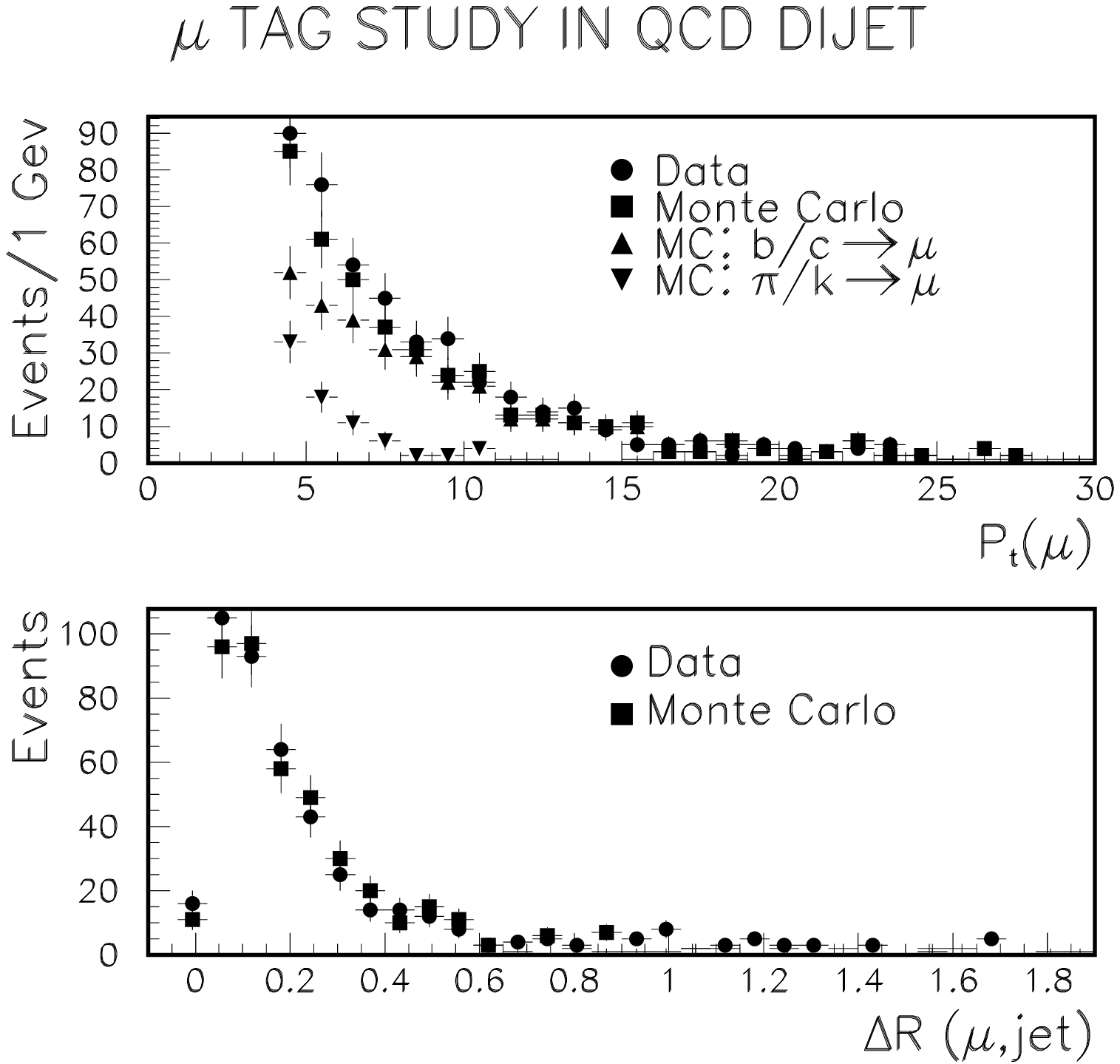,width=3.0in,height=2.5in}}
\caption{Comparison of data and Monte Carlo predictions for QCD dijet events
containing a muon}
\label{fig1}
\end{figure}

In order to test our understanding of muon and jet reconstruction efficiencies,
we look for soft muons in a QCD dijet sample of events. \Fref{fig1}  shows  the
\pet\ spectrum of the muons. Also shown are the Monte Carlo  \cite{geant}
calculations of the contributions from muons resulting from $\pi$ and $K$ decay
and $b$ and $c$ quark decay. The sum of these two contributions reproduces the
data well for \pet\ $>$ 4 \gc\ . Also shown in the figure is the separation
$\Delta R$ in  $\eta \times \phi$ space of the muon and the nearest jet. The
Monte Carlo again reproduces this distribution well.

\begin{figure}
\centerline{\psfig{figure=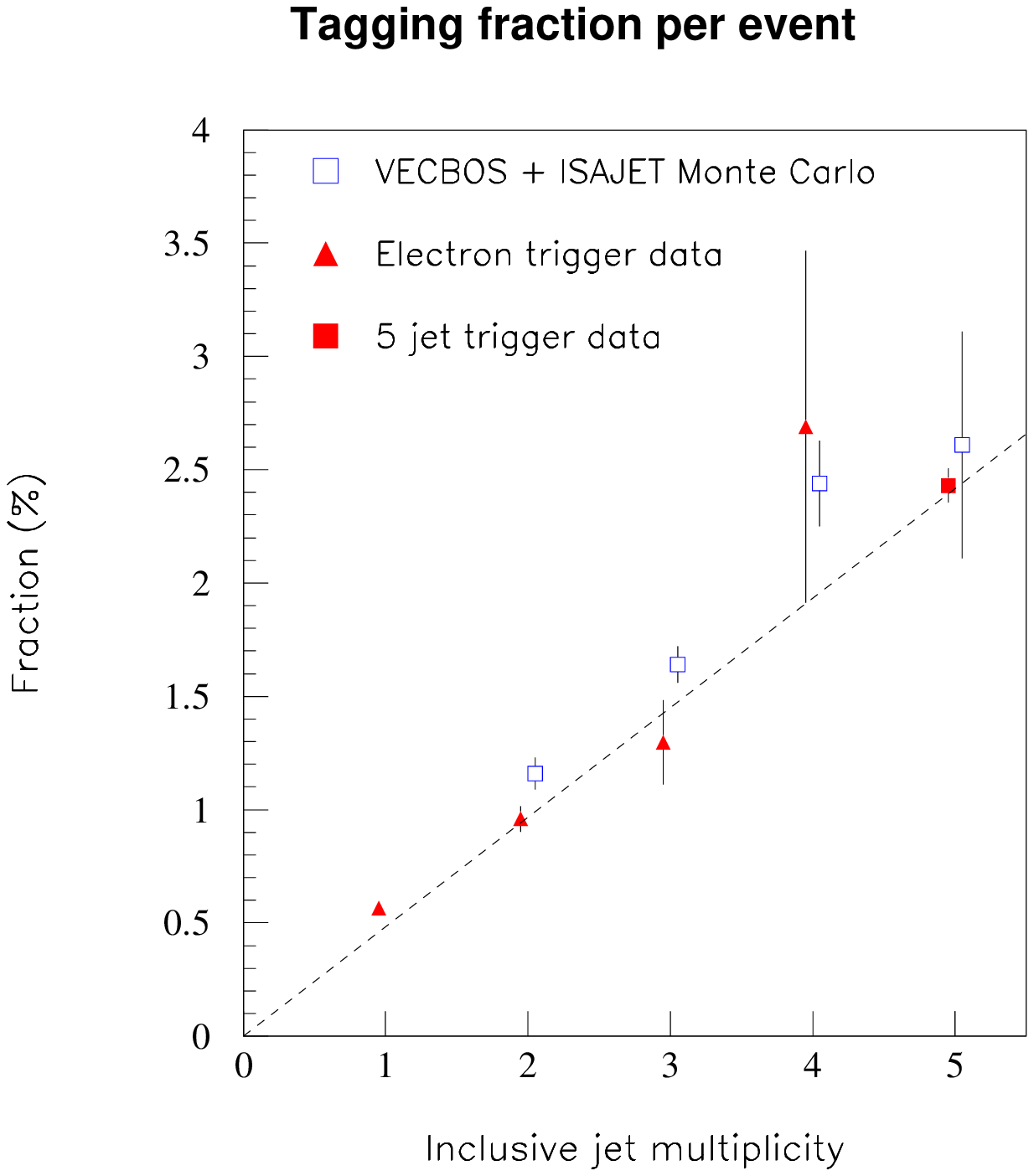,width=3.0in,height=2.5in}}
\caption{Fraction of events containing muons as a function of the inclusive
jet multiplicity}
\label{fig2}
\end{figure}

There are two main sources of background to the channel e+ jets + soft $\mu$
tag
from \tbt\ production. The first  is from W+Jets production where some of the
jets  result from the fragmentation of $b$ and $c$ quarks. The second is from
QCD multi-jet production containing $b$ or $c$ quarks where one of the jets
fakes an electron and the \met\ is produced primarily by detector resolution.
In
each case we assume that the probability for a jet to emit a detectable muon is
independent of the process producing the jet and is  a function of the \et\ of
the jet. The source of the muon may be $b$ or $c$ quark decay, $\pi$ or $K$
decay or fake $\mu$'s due to reconstructing random hits in the muon chambers.
We justify this assumption by examining the fraction of events that contain a
jet tagged by a muon as a function of the inclusive jet  multiplicity (defined
as  multiplicity $\geq$ a given number of jets) for three different sets of
events; for data triggered on a single high \et\  ($\geq$ 20 \g) electron, for
QCD 5 jet data and for VECBOS \cite{vecbos} Monte Carlo  that describes W+Jets
production that has been put through the Isajet \cite{isajet} shower
fragmenter.
The results are shown in  \fref{fig2}.  The muon tagging fraction is linearly
proportional to  the jet multiplicity. The probability for a jet to emit a
detectable muon seems to be $\sim$ 0.5 \%, justifying the above assumption.
\subsection{Definition of the QCD Fake sample}
\begin{figure}
\centerline{\psfig{figure=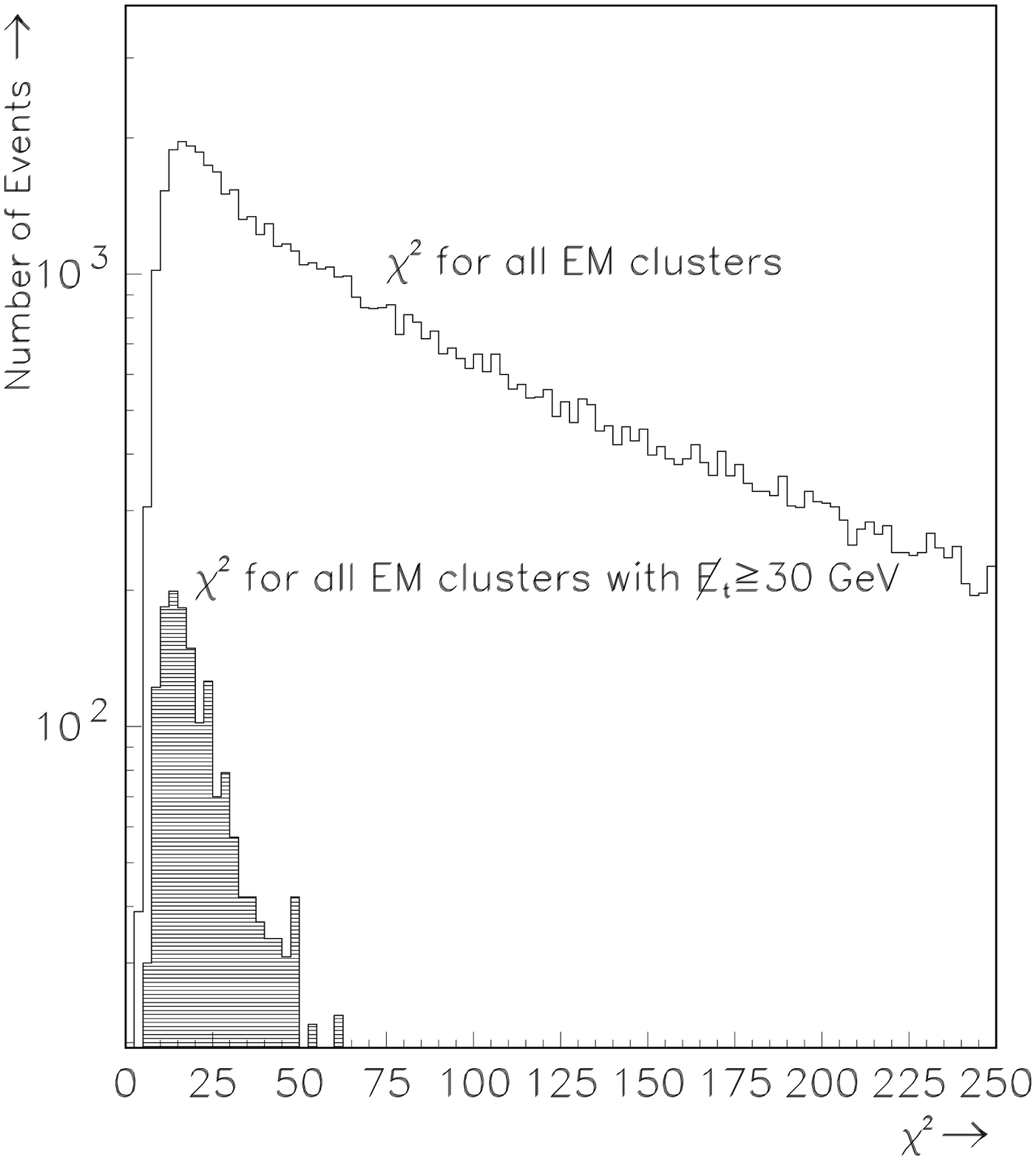,width=3.0in,height=2.5in}}
\caption{H matrix $\chi^2$ distribution for all EM clusters and for
EM clusters with \met\ $>$ 30 \g\ (shaded)}
\label{fig3}
\end{figure}

In order to extract the tagging fraction function from data, we first isolate
a sample of events which possess a fake electron but which in all other
respects
resemble the electron + jets event sample under study. Our electron
identification algorithm uses a Fisher $\chi ^2$ discriminant variable based on
41 quantities describing the energy deposition of the electron in the
calorimeter. The $\chi ^2$ variable is described as follows.
\begin{eqnarray*}
 E_{ij} = <x_ix_j> - <x_i><x_j> \\
 \chi^2 = \Sigma_{ij} (x_i-<x_i>)H_{ij}(x_j-<x_j>)
\end{eqnarray*}
where the covariance matrix E and its inverse H matrix are defined in terms of
the 41 dimensional vector $x$, which consists of three longitudinal energy
fractions, 36 transverse energy fractions at shower maximum, log(Energy of
cluster) and the position of the vertex along the beam direction. The angular
brackets $<>$ in the above equations signify averages over events. We employ a
different H matrix for each of the 37 towers in pseudo-rapidity for either half
of the calorimeter. \Fref{fig3} shows the $\chi^2$ distribution for all
electromagnetic clusters  with \et\ $>$ 20 \g\ and for those which have  \met\
$>$ 30 \g\ . These latter are dominated by genuine electrons from W's and have
a
much narrower  $\chi^2$ distribution. In  defining good electrons, we
demand that the $\chi^2$ is less than 100. In addition, we define a track
match significance parameter as the error weighted impact parameter between the
central detector track and the cluster centroid in the azimuthal and beam
directions. We demand  a central detector track that passes close  to the
shower
centroid  with a track match significance of less than 5 for good electrons.
Since we are interested in isolated electrons, we demand the isolation fraction
to be less than 0.1. The isolation fraction is defined as
\begin{eqnarray*}
\frac{(Total\: Energy \:in \:0.4 \:cone\: - \:EM \:Energy\: in\: 0.2 \:cone)}
                          {EM\: energy\: in \:0.2 \:cone}
\end{eqnarray*}
where the cone size is in $\Delta R $
space. We define a fake electron as any EM cluster that fails the good electron
criteria and the QCD fake sample as those triggers that have electromagnetic
clusters with \et\
$>$ 20 GeV and fail the good electron criteria with no requirement on \met\ .
\subsection{Determination of tagging fraction function}
\begin{figure}
\centerline{\psfig{figure=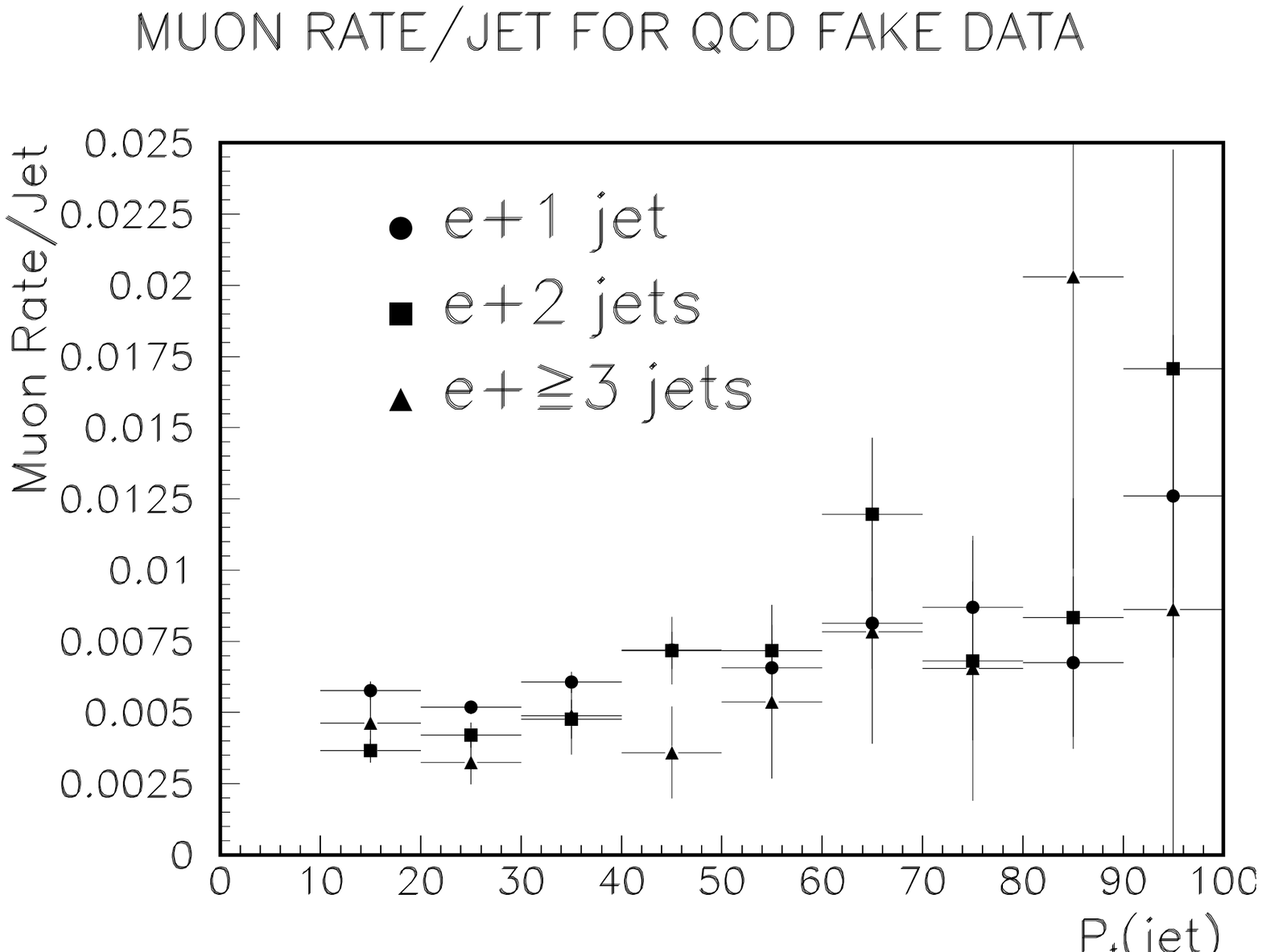,width=3.0in,height=2.5in}}
\caption{Jet tagging fraction vs \et\ of jet for QCD fake events}
\label{fig4}
\end{figure}

\begin{figure}
\centerline{\psfig{figure=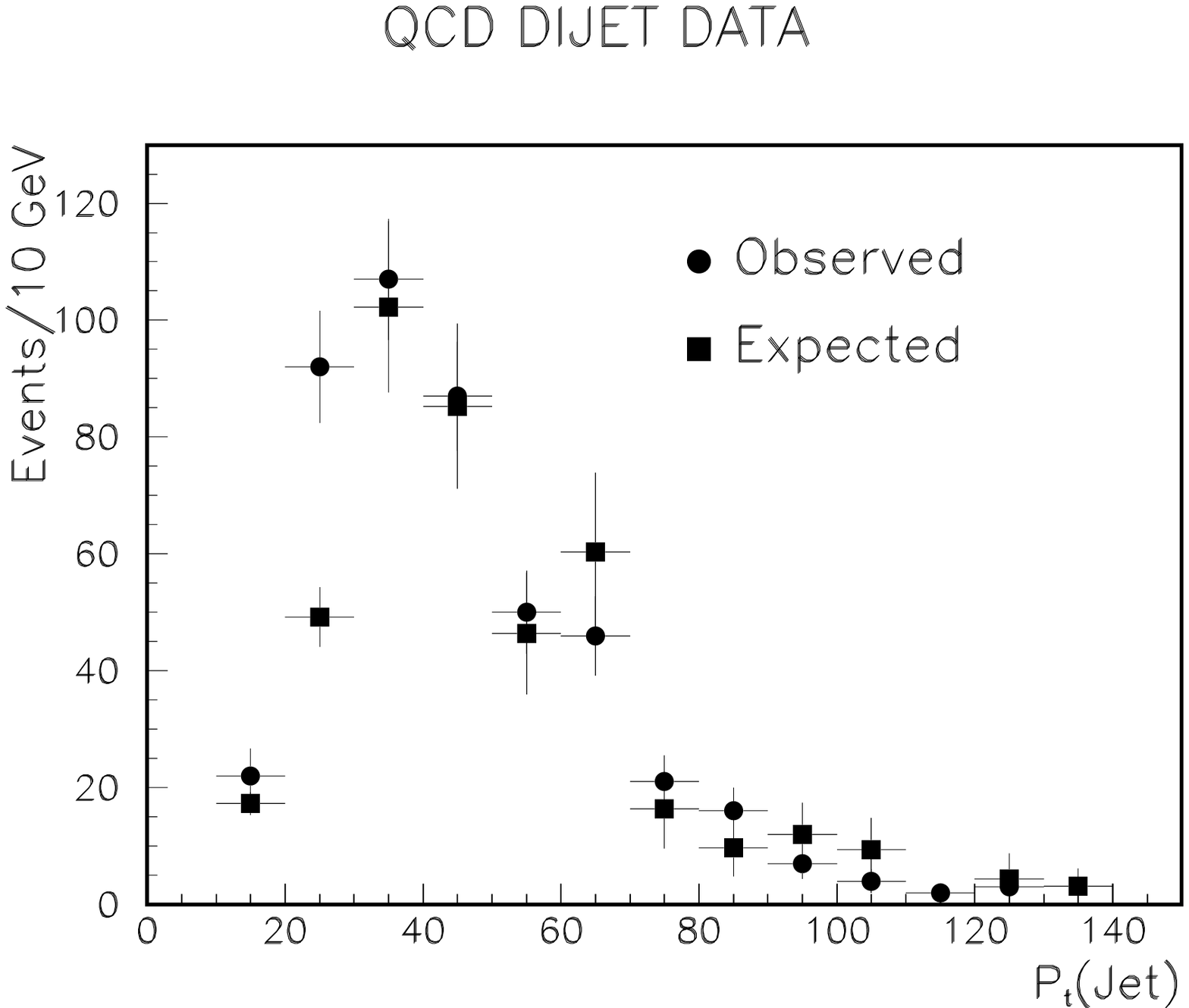,width=3.0in,height=2.5in}}
\caption{\et\ spectrum of  of tagged jets in QCD dijets, data and prediction}
\label{fig5}
\end{figure}

\begin{figure}
\centerline{\psfig{figure=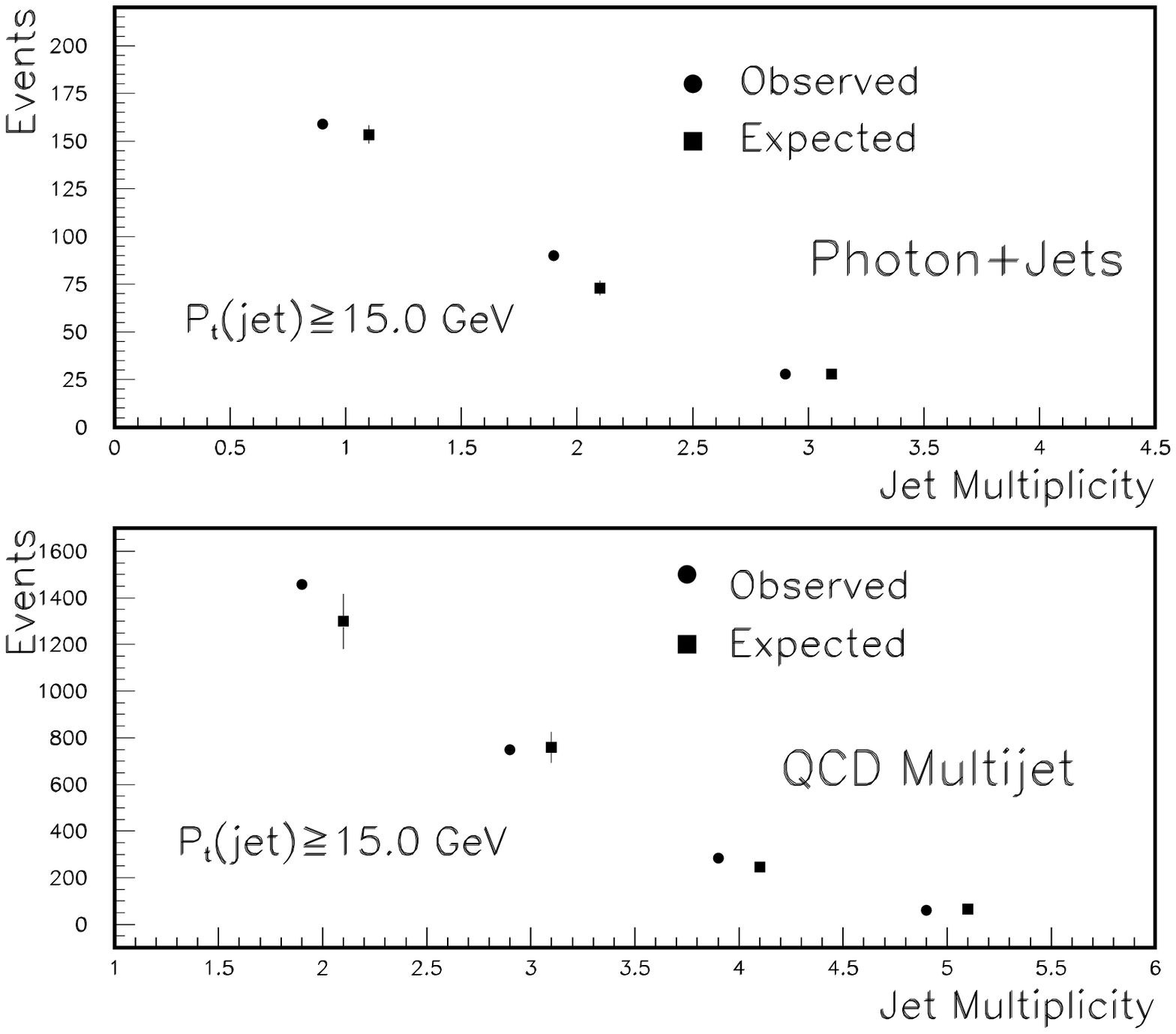,width=3.0in,height=2.5in}}
\caption{Jet multiplicity distribution of tagged events for  ``photon" + jets
and QCD multijets}
\label{fig6}
\end{figure}

We now use the QCD fake sample as a source of jets and determine the fraction
of
jets that have muons as a function of \et\ of the jet {\it and} jet
multiplicity. We require muons to have \pet\ $>$ 4.0 \gc\ and $|\eta|<$ 1.7. We
demand that the muon be non-isolated if its \pet\ is greater than 12 \gc\ .
This
selection makes this event sample exclusive of the $e\mu$ sample in reference
\cite{steve}.
\Fref{fig4} shows the jet tagging fraction as a function of \et\ for the QCD
fake sample for jet multiplicities of 1,2 and $\geq$ 3 jets. We now assume that
this tagging function, determined as a function of \et\ and multiplicity is
universal. As a cross check of this hypothesis, we test this on QCD dijet data.
\Fref{fig5} shows the \et\ spectrum of jets with tagged muons in QCD dijets and
the spectrum that is predicted assuming the above tagging functions. There is
seen to be good agreement between prediction and data, which gives us
confidence
in the hypothesis. As a further cross check, we examine the jet multiplicity
distribution of tagged jets in ``photon" + jets candidates and QCD multijets. A
``photon" is an electromagnetic cluster which passes all the good electron
criteria except that it has no central detector track. \Fref{fig6} shows the
distribution of jet multiplicity for these two sets of data and the prediction
using the tagging fraction function. Again there is seen to be good agreement.
In order to calculate the $\mu$ tag background in W+jets due to the presence of
$b$ and $c$ quarks associated with W production, we apply the tagging fraction
functions to the W+ Jets sample.
\subsection{Calculation of the W+ jets + $\mu$ tag background}
\begin{figure}
\centerline{\psfig{figure=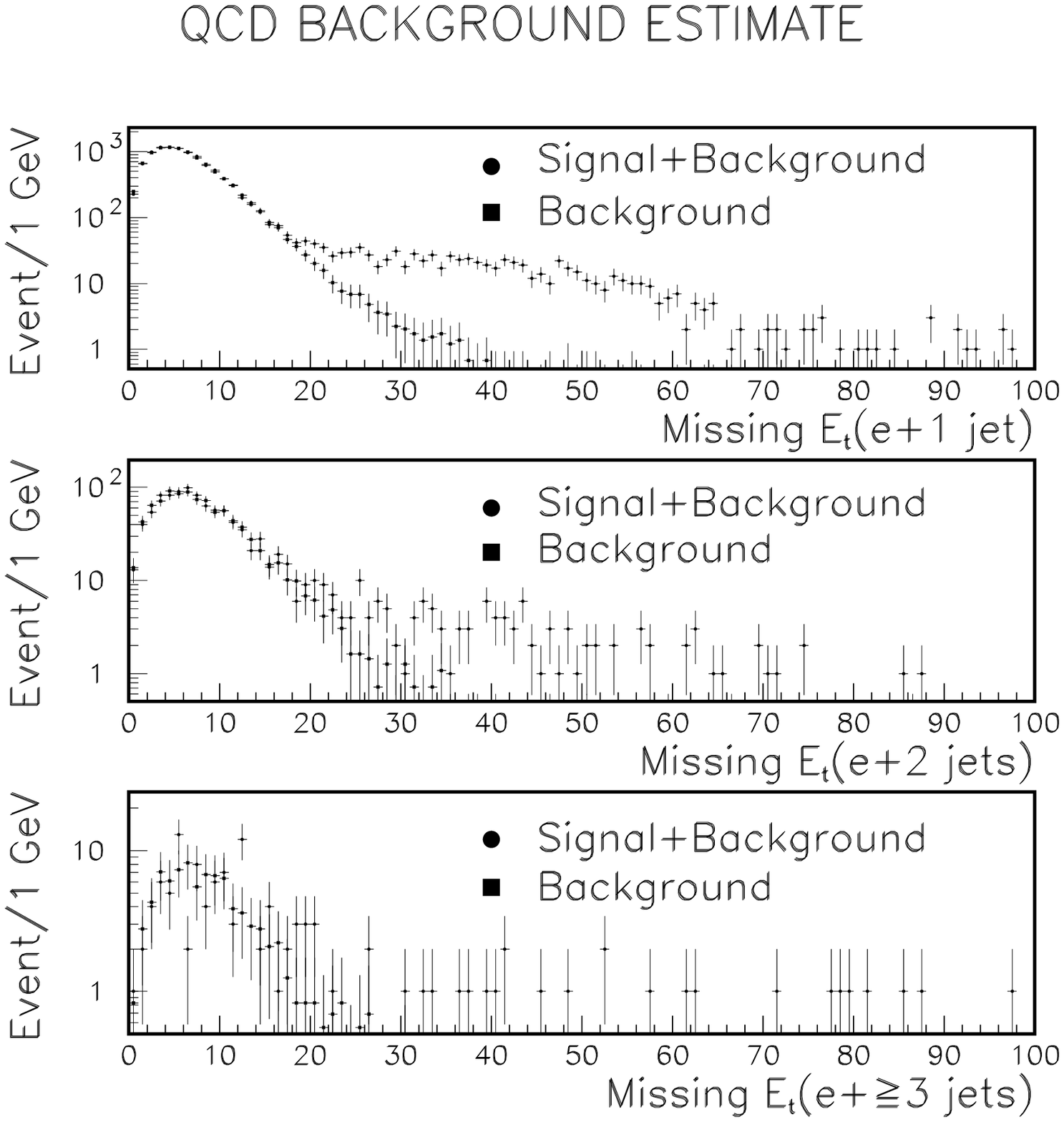,width=3.0in,height=4.5in}}
\caption{\met\ distribution of W + jet data and
QCD fake background normalized to data for different jet multiplicities}
\label{fig7}
\end{figure}

\Fref{fig7} shows the
\met\ distribution of W+Jet data. The QCD fake background is normalized to the
data for \met\ $<$ 15 \g\ . We now subtract the QCD fake background from the
W+jets data (\met\ $>$ 20 \g) to obtain the total amount of W+Jets production.
We apply the tagging fraction function to the amount of signal thus obtained.
We
handle the QCD fake contribution to tagged events separately, since the QCD
fakes are at lower \met\ and the presence of the muon affects the \met\
distribution sufficiently to warrant a separate calculation.
\subsection{Calculation of the QCD fake $\mu$ tag background}
\begin{figure}
\centerline{\psfig{figure=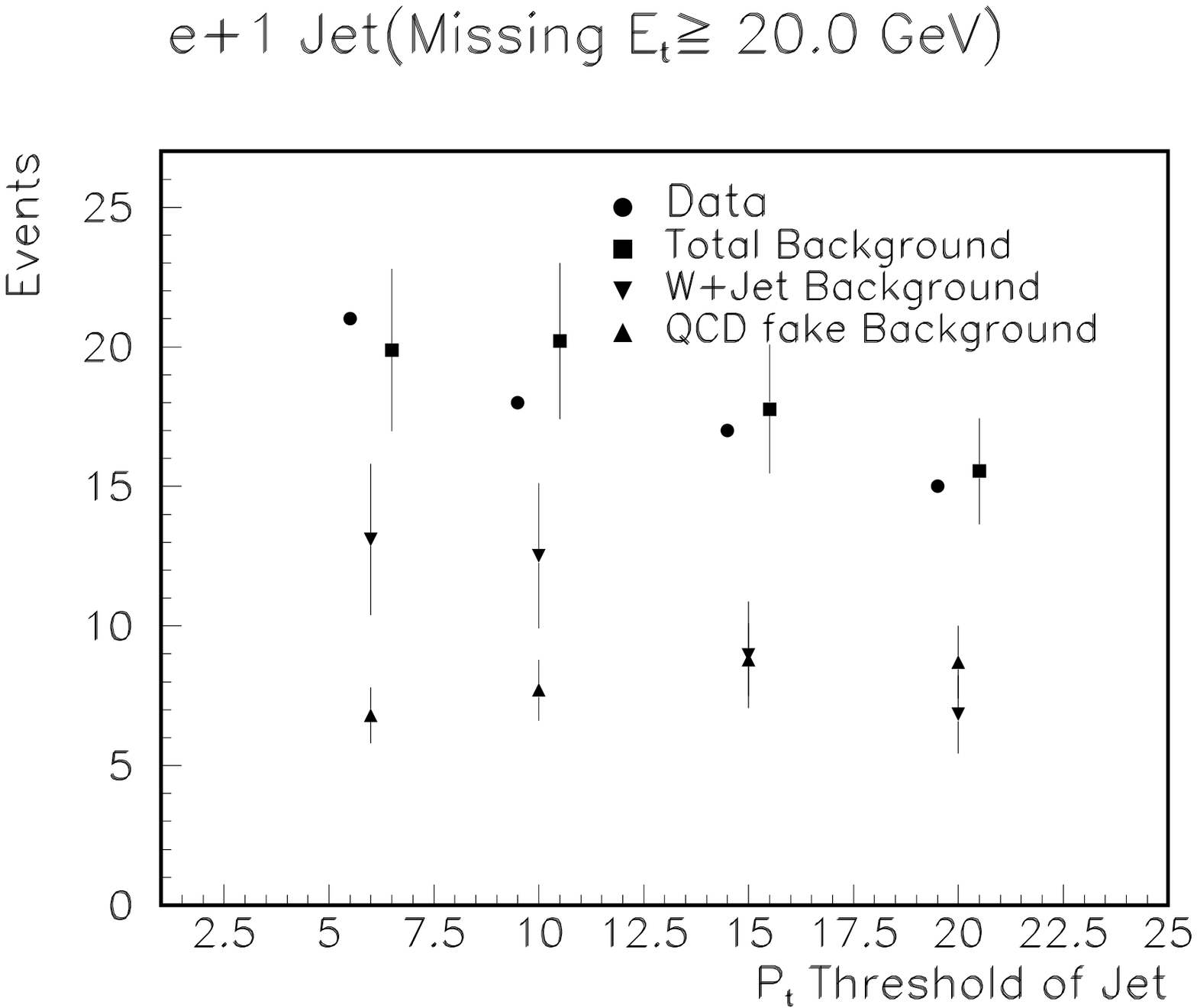,width=3.0in,height=2.5in}}
\caption{Comparison of background predictions and data for electron + 1 jet
events with \met\ $>$ 20 \g}
\label{fig8}
\end{figure}

Since we have normalized the QCD fakes to the W+jets signal for  \met\ $<$ 15
\g\ , we estimate the QCD fake background by normalizing  the tagged muon
events in the QCD fake sample with \met\ $>$ 20 \g\ , by the same factor. We
now attempt one further cross check, by comparing the W + 1 Jet data (with
\met\ $>$ 20 \g) with the background predictions. Very little top is expected
with 1 jet only. \Fref{fig8} shows the comparison of background predictions
with data, as a function of \et\ of the jet. The agreement between predicted
and
observed  values is good.
\section{Summary of cuts and the surviving signal and background}
\begin{figure}
\centerline{\psfig{figure=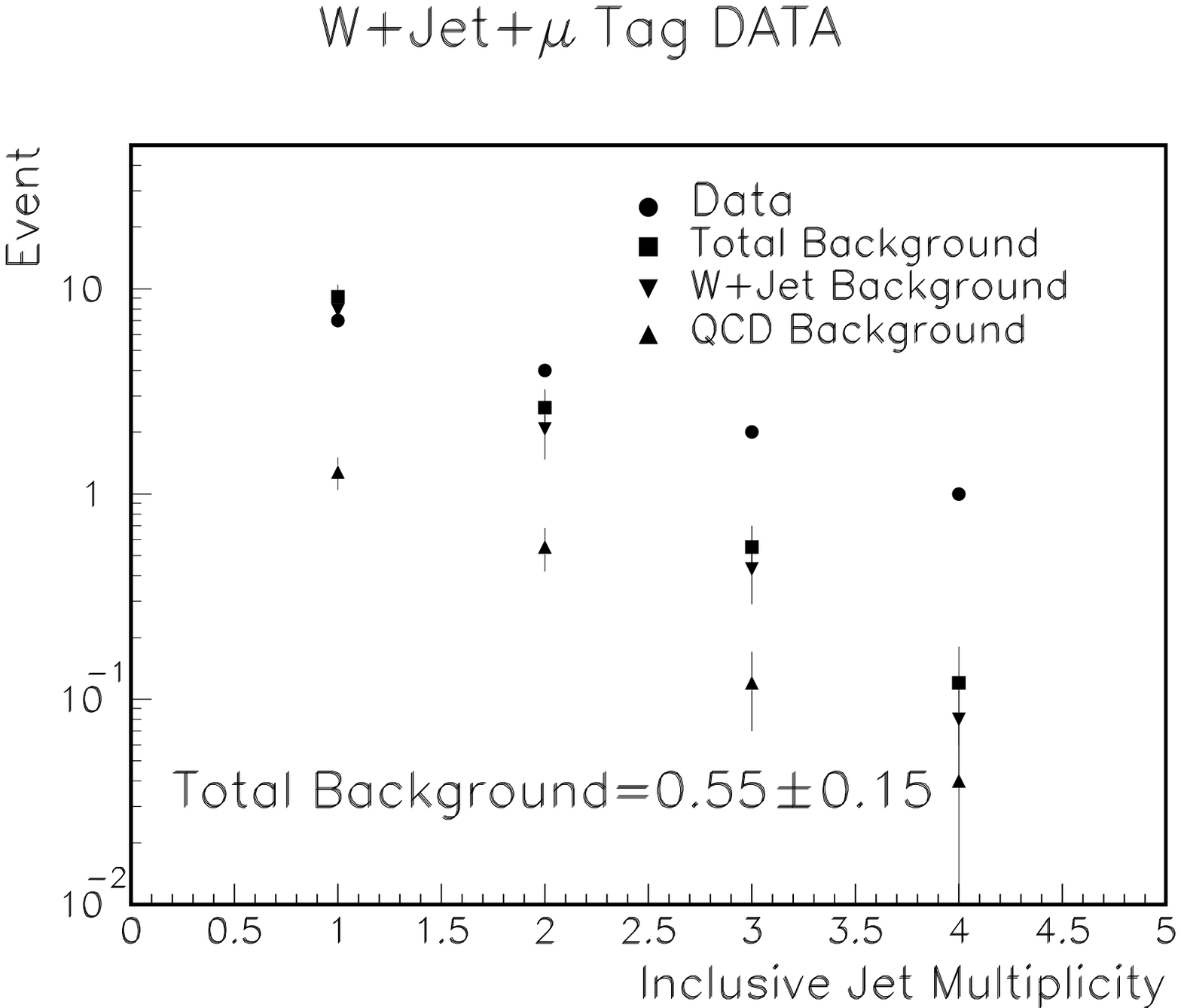,width=3.0in,height=2.5in}}
\caption{Comparison of background predictions and data as a function of
inclusive jet multiplicity.}
\label{fig9}
\end{figure}

\Tref{tabl1} shows the summary of the cuts used, the surviving number
of events and background estimates as well as the expectation
from top production at various masses \cite{laenen}.
The $\Delta\phi (\mu, \met\ )$ cut is introduced to take into account the
correlation between \met\ and the muon \pet\ for QCD fake events.
Two events survive the cuts
described with a total  expected background of 0.55 $\pm$ 0.15 events.
\Fref{fig9} compares the data and background predictions as a function of
inclusive jet multiplicity.
\begin{table}
\Table{|c|c|}
{Particle type & Cuts \\
\hline
Electron & H-Matrix $\chi^2  <  100$ \\
         & Track match signif. $ < $ 5.0 \\
         & $|\eta| < 2.0 $ \\
         & $ E_T > 20\: \g\ $ \\
         & dE/dx minimum ionizing \\
\hline
Muon     &  $|\eta|< $ 1.7 \\
         &  \pet\ $>$ 4 \gc\ \\
         &  non-isolated muon or \\
         & \pet\  $<$ 12 \gc\ \\
\hline
\met\    & $>$ 20 \g\ \\
         & $\Delta\phi (\mu, \met\ ) > 25^o  $ \\
         & if \et\ $<$ 35 \g\ \\
\hline
Jets     & $\geq$ 3 jets \et\ $>$ 20 \g\ \\
\hline
Data     & Events \\
         &    2   \\
\hline
Background & Events \\
W + jets   & 0.43 $\pm$ 0.14 \\
QCD fakes  & 0.12 $\pm$ 0.05 \\
Total      & 0.55 $\pm$ 0.15 \\
\hline
Top mass \gsq\  & Expected events \\
140            & 1.3 $\pm$ .4 \\
160            & 1.0 $\pm$ .2 \\
180            & 0.6 $\pm$ .2 \\}
\caption{ Summary of cuts, data, background and top yields \label{tabl1}}
\end{table}
\section{Combined top cross section and conclusions}
\begin{figure}
\centerline{\psfig{figure=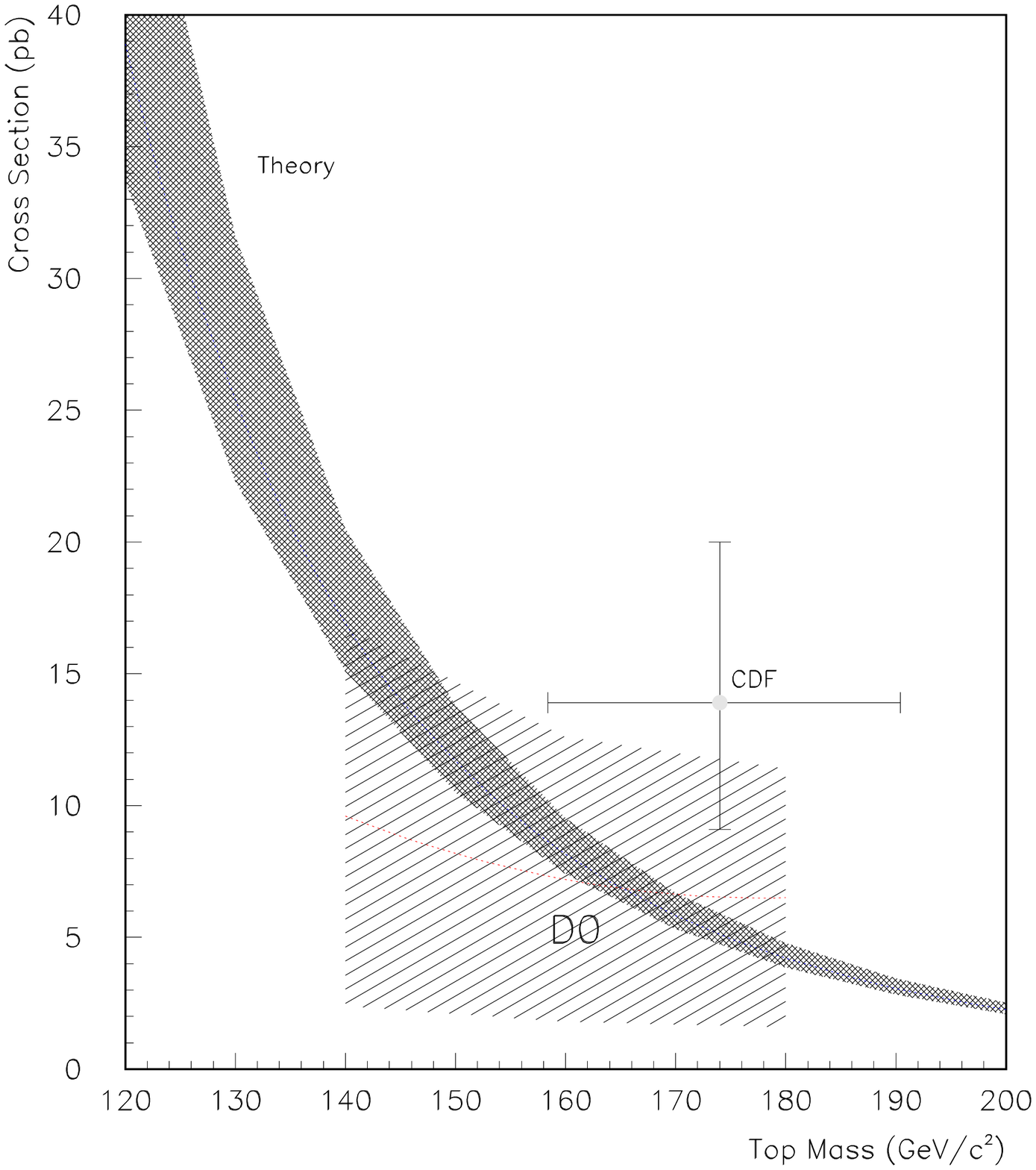,width=3.0in,height=3.5in}}
\caption{D\O\ top cross section results compared with theoretical predictions
and CDF}
\label{fig10}
\end{figure}

We now combine the results of various D\O\ top searches \cite{steve,serban}
reported at this conference with the tagged muon results reported here to
obtain
a top cross section and error. \Tref{tabl2} summarizes the numbers  reported in
all the channels. \Fref{fig10} gives the D\O\ results as a function of top mass
compared to theoretical predictions \cite{laenen} and the recently reported CDF
result \cite{cdf}. Expressed in terms of top production cross section, D\O\
obtains cross sections of  9.6 $\pm$ 7.2 pb, 7.2 $\pm$ 5.4 pb, 6.5 $\pm$ 4.8 pb
for top masses of 140, 160 and 180 \gsq. This assumes that top quark decays
with standard model decay modes. This is consistent both with a null result as
well  as the published CDF result. The D\O\ $\mu$ + jets with $\mu$ tag
analysis
is still in progress. D\O\ is also pursuing multivariate analyses with an aim
to
increase our signal acceptance for a given background rejection as well as mass
analyses of the lepton + jets candidates. With the increased statistics of the
current Tevatron run, we should be able to considerably increase our discovery
limit for the top quark very shortly. The results presented here should be
regarded as preliminary.
\begin{table*}
\Table{|cc|c|c|c|c|c|c|c|}
{\hline
\multicolumn{2}{|c|}{$m_t$ [GeV/$\rm c^2$]} &
\multicolumn{1}{c|}{$e\mu$} &
\multicolumn{1}{c|}{$ee$~~} &
\multicolumn{1}{c|}{$\mu\mu$~~~} &
\multicolumn{1}{c|}{$e + \rm jets$~} &
\multicolumn{1}{c|}{$\mu+\rm jets$~ } &
\multicolumn{1}{c|}{$e + \rm jets (\mu)$} &
\multicolumn{1}{c|}{ALL} \\ \hline \hline
      & $\varepsilon \times B (\%) $ & $.32 \pm .06$ & $.18\pm .02$
      & $.11\pm .02$ & $1.2\pm 0.3$ & $.8\pm 0.2$ &  $0.6\pm 0.2$ & \\
            \cline{2-9}
 ~140 & $\langle N \rangle$             & $.72 \pm .16$ & $.41\pm .07$
      & $.24\pm .05$ & $2.8\pm 0.7$ & $1.3 \pm 0.4$  &   $1.3 \pm 0.4$
      & $6.7\pm 1.2$ \\ \hline
      & $\varepsilon \times B (\%) $ & $.36 \pm .07$ & $.20\pm .03$
      & $.11\pm .01$ & $1.6\pm 0.4$ & $1.1\pm 0.3$ &  $0.9\pm 0.2$ & \\
            \cline{2-9}
 ~160 & $\langle N \rangle$             & $.40 \pm .09$ & $.22\pm .04$
      & $.12\pm .02$ & $1.8\pm 0.5$ & $0.9 \pm 0.3$  &   $1.0 \pm 0.2$
      & $4.4\pm 0.7$ \\ \hline
      & $\varepsilon \times B (\%) $ & $.41 \pm .07$ & $.21\pm .03$
      & $.11\pm .01$ & $1.7\pm 0.4$ & $1.2\pm 0.3$ & $1.1\pm 0.2$ & \\
            \cline{2-9}
 ~180 & $\langle N \rangle$             & $.23 \pm .05$ & $.12\pm .02$
      & $.06\pm .01$ & $1.0\pm 0.2$ & $0.5 \pm 0.2$  &   $0.6 \pm 0.2$
      & $2.5\pm 0.4$ \\ \hline \hline
\multicolumn{2}{|c|}{Background}    & $.27 \pm .09$ & $.16 \pm .07$ &
$.33 \pm .06$ & $1.2 \pm 0.7$ & $0.6 \pm 0.5$ & $0.6 \pm 0.2$  &
$3.2 \pm 1.1$ \\ \hline \hline
\multicolumn{2}{|c|}
{$\int {\cal L}dt \ [\rm pb^{-1}]$} & $13.5 \pm 1.6$ & $13.5 \pm 1.6$
& $9.8 \pm 1.2$ & $13.5 \pm 1.6$ & $9.8 \pm 1.2$ & $13.5 \pm 1.6$ & \\ \hline
\multicolumn{2}{|c|}{Data} & \multicolumn{1}{c|}{ 1} &
\multicolumn{1}{c|}{0~~} & \multicolumn{1}{c|}{0~} &
\multicolumn{1}{c|}{2~~} & \multicolumn{1}{c|}{2~~} &
\multicolumn{1}{c|}{2~~~} &
\multicolumn{1}{c|}{7~} \\ \hline}
%\normalsize
%\narrowtext
\caption{Efficiency $\times$ branching fraction ($\varepsilon \times B$),
expected number of events ($\langle N \rangle$) for
signal
and background sources for the observed integrated luminosity
($\int {\cal L}dt$), and number of events observed in the data. \label{tabl2}}
%\caption{ Summary of results from dilepton, lepton + jets and e+jets with
%%$\mu$
%tag results \label{tabl2}}
\end{table*}

\end{document}